\documentclass[onecolumn]{aastex}
\usepackage{spr-astr-addons}
\usepackage{graphicx}
\usepackage{txfonts}
\usepackage{xcolor}
\usepackage{tikz}
\usepackage{rotating}
\usepackage{lscape}
\usepackage{adjustbox}
\usepackage[version=4]{mhchem}
\usepackage[normalem]{ulem}
\usepackage{float}
\usepackage{natbib}
\pdfoutput=1 
\usepackage{amsmath,amstext,amssymb}
\usepackage[T1]{fontenc}
\usepackage[colorlinks=true, allcolors=blue]{hyperref}

\RequirePackage{color}
\begin{document}
	
\title{Multi-wavelength observation of MAXI J1348--630 during the outburst in 2019}

	\shortauthors{Mandal et al.}

	\author{Manoj Mandal\altaffilmark{1}} \author{Debasish Saha\altaffilmark{2}} \author{Sabyasachi Pal\altaffilmark{1}} \and \author{Arijit Manna\altaffilmark{1}}
	
	\email{sabya.pal@gmail.com}
	
	\altaffiltext{1}{Midnapore City College, Paschim
		Medinipur, West Bengal, India 721129 \\email: {sabya.pal@gmail.com}}
	
	\altaffiltext{2}{Department of Physics, Indian Institute of Science Education and Research, Bhopal, Madhya Pradesh, India, 462066}

	\begin{abstract}
We study the multi-wavelength spectral properties of the black hole X-ray binary MAXI J1348--630 using quasi-simultaneous \textit{ALMA}, \textit{NICER}, and \textit{Swift} observations during the decay phase of the January 2019 outburst. In millimeter wavelengths, radio continuum emissions in the frequency range of 89.56--351.44~GHz are measured. We found that the flux densities at millimeter wavelength varied between 12.18 mJy and 18.47 mJy with spectral index ($\alpha $) of $0.28\pm 0.02$. The broadband spectrum suggests that
the source was accompanied by weak synchrotron emission from the compact jets. Broadband spectral study indicates that MAXI J1348--630 falls in the regime of ``radio-quiet'' during the decay phase of the outburst. The \textit{NICER} spectrum is fitted by a combined model of disk blackbody component
(\textit{diskbb}) along with a comptonization component (\textit{simpl}) which explains the power-law continuum caused by the thermal Comptonisation of soft disk photons in a hot gas of electrons. The \textit{NICER} spectrum is dominated by the comptonised components during the decay phase of the outburst close to the hard state of the source. We have investigated the correlation between X-ray and radio luminosity using quasi-simultaneous \textit{ALMA} and \textit{NICER} data to understand the source nature by locating the source in the $L_{X}$-$L_{R}$ diagram. The correlation study of radio/X-ray luminosity suggests that MAXI J1348--630 did not follow the well-known track for black
holes and it is a new member of a restricted group of sources.
\end{abstract}


\section{Introduction}
\label{sec:intro} 
Black hole transients are known to accrete matter from companion stars, which form an accretion disk around them, and the emission of X-rays is observed during the event. During an outburst, transitions between different spectral states are observed with various timing and spectral properties \citep{Ho05}. Black hole transients spend a large duration of time (several months to years) in quiescence, and during outbursts, emissions in different wavelengths are observed \citep{Ca11}, and the source luminosity increases by several orders of magnitude in all wavelengths of the electromagnetic spectrum \citep{Re06}. Outbursts of black hole X-ray binaries (BHXRBs) are characterized by their evolution through different spectral states \citep{Re06, Be09}. At the beginning of the outburst, they remain in the hard state (HS), during which comptonized hard photons dominate the spectra. As luminosity increases, they usually make a transition to the soft state (SS). In SS, very strong emissions of soft X-rays from the accretion disk are observed. During this transition from HS to SS, BHXRBs sometimes exhibit intermediate states, which are known as hard intermediate states (HIMS) and soft intermediate states (SIMS) \citep{Ho05}. In most cases, BHXRBs are in the SS when they reach the maximum luminosity, and similar state
transitions are observed in the decay phase of the outburst until they
return to the HS \citep{Ho05, Va06, Be11, Mo11}. The complete evolution
generally creates a ``$q$''--shaped track in the hardness-intensity diagram
(HID) \citep{Ho05, Re06, Du11}. The spectral parameters are used to probe
the emission mechanism and different spectral states of a black hole. During
SS, spectra are steep, with a typical spectral index of $(\Gamma ) \geq 2.5$ \citep{Re06}. Accretion disks are relatively hot, with a disk temperature of $kT \approx 1$ keV \citep{Sh73, Du11}, and thermal X-rays are emitted from the innermost regions of the accretion disks. Spectra becomes hard with $\Gamma \approx 1.6$ in the HS, and the disk becomes relatively cooler \citep{Ma08, Mi08, Ta08}. Multi-wavelength observations help to understand emission mechanisms, and accretion phenomena of a source during the outburst.

BHXRBs show emission from radio to IR frequencies, which is attributed to synchrotron emission from compact jets \citep{Fe06, Ga10}. Positive correlations between radio and X-ray luminosities are generally found \citep{Ga03, Ca21} which indicates a coupling nature between the emissions from jets and accretion disks \citep{Ya01, Co03}. The BHXRB transient MAXI J1348--630 was discovered during a giant outburst in January 2019 \citep{Ya19} and followed by a subsequent outburst in June 2019 \citep{Ne19}. During the January 2019 outburst, the source reached a peak flux of $\sim $4 Crab as measured by \textit{MAXI}/GSC in 2--20~keV \citep{To20}. The optical counterpart of MAXI J1348--630 was detected by
\citet{De19} during the January 2019 outburst. Previous observations suggested that the source is a black hole candidate \citep{Sa19, To20, Zh20}. The mass of the black hole was estimated to be $\sim $7$M_{\odot}$ and \citet{Ch21} measured the source distance to be $\sim $2.2 kpc by studying the absorption line of HI observed by the Australian Square Kilometre Array Pathfinder (\textit{ASKAP}) and \textit{MeerKAT}. Different types of quasi-periodic oscillations (QPOs; types
A, B, and C) were found during the January 2019 outburst \citep{Be20, Ja20, Zh20}.

Using \textit{Insight-HXMT}, \citet{We21} conducted a detailed time-lag study of MAXI J1348--630 from which it was concluded that the observed time lag between the radiations of the accretion disk and corona causes the hysteresis feature and the ``q''-diagram. \citet{Ji22} investigated the black hole spin parameter and the inclination of the accretion disk. The evolution of outburst and different timing properties of MAXI J1348--630 was studied using multiple \textit{NICER} observations \citep{Zh20}. To track the evolution of the jets, simultaneous X-ray (\textit{MAXI} and \textit{Swift}/XRT) and radio (\textit{MeerKAT} and \textit{ATCA}) observations were used \citep{Car21}.

Earlier, several radio/X-ray observations were performed to understand the multi-wavelength behaviour of MAXI J1348--630 during the 2019 outburst. The radio flux density at 1.28~GHz was 520.3 $\pm $ 5.0 mJy on MJD 58523.21 as observed by \textit{MeerKAT} \citep{Ca19}. In another radio observation by the \textit{Murchison Widefield Array (MWA)} telescope on MJD 58523.79, flux densities at 154~MHz and 216~MHz were 301 $\pm $ 21 mJy and 362 $ \pm $ 22 mJy respectively \citep{Ch19}. The quasi-simultaneous X-ray flux was 1 $\times $ 10$^{-8}$ erg cm$^{-2}$ s$^{-1}$ as observed by \textit{Swift}/XRT (1--10~keV) on MJD 58523.44.

In this paper, we study the multi-wavelength spectral properties of MAXI J1348--630 using fluxes in radio, optical, ultra-violet, and X-ray wavelengths during the decay phase of the January 2019 outburst. We also study the correlation of radio/X-ray luminosities to understand the nature of the source using quasi-simultaneous \textit{ALMA} and \textit{NICER} observation. The paper is organized in the following manner: Observation and data analysis methods of different instruments are discussed in Sect.~\ref{sec:obs}, and results are summarised in Sect.~\ref{sec:result}. The discussion and the conclusions are summarised in Sects.~\ref{sec:discussions} and \ref{sec:conclusion}, respectively.
	
	
	\begin{table*}
		\centering
		\caption{Log of {\it ALMA}, {\it NICER}  and {\it Swift} observations} 
		\label{tab:log}
		\begin{tabular}{lcccccc}
			\hline
			Instruments	& Obs. ID	& Band/Filter		            & Date		    & Start Time	& MJD		& Exposure \\
			& 	& 		            & 	    & 	& 	& (ks) \\
			\hline
			& 2018.1.01034.T	& Band-3 ~~(89.56--105.43 GHz)	& 2019-05-02 	& 04:15:00 	    & 58605.18	& ~0.302	\\
			{\it ALMA}	& 2018.1.01034.T	& Band-4 (137.06--152.93 GHz)	& 2019-05-02 	& 04:34:00 	    & 58605.19	& ~0.302	\\
			& 2018.1.01034.T	& Band-6 (223.07--105.43 GHz)	& 2019-05-02 	& 03:50:00 	    & 58605.16	& ~0.423	\\
			& 2018.1.01034.T	& Band-7 (335.56--351.44 GHz)	& 2019-05-02 	& 03:06:00 	    & 58605.13	& ~1.149	\\
			\hline
			{\it NICER} & 2200530128	&  0.5--10 keV  & 2019-05-02 	& 07:08:39	    & 58605.30	& 3.869 	\\
			\hline
			{\it Swift} (UVOT) & 00011107032	& uvm2 ($\lambda=2246$ \AA)     & 2019-05-03    & 09:25:35      & 58606.39  & 0.526	\\
			\hline
		\end{tabular}
	\end{table*}
	
	\begin{figure*}[t]
		\centering{
			\includegraphics[width=15cm]{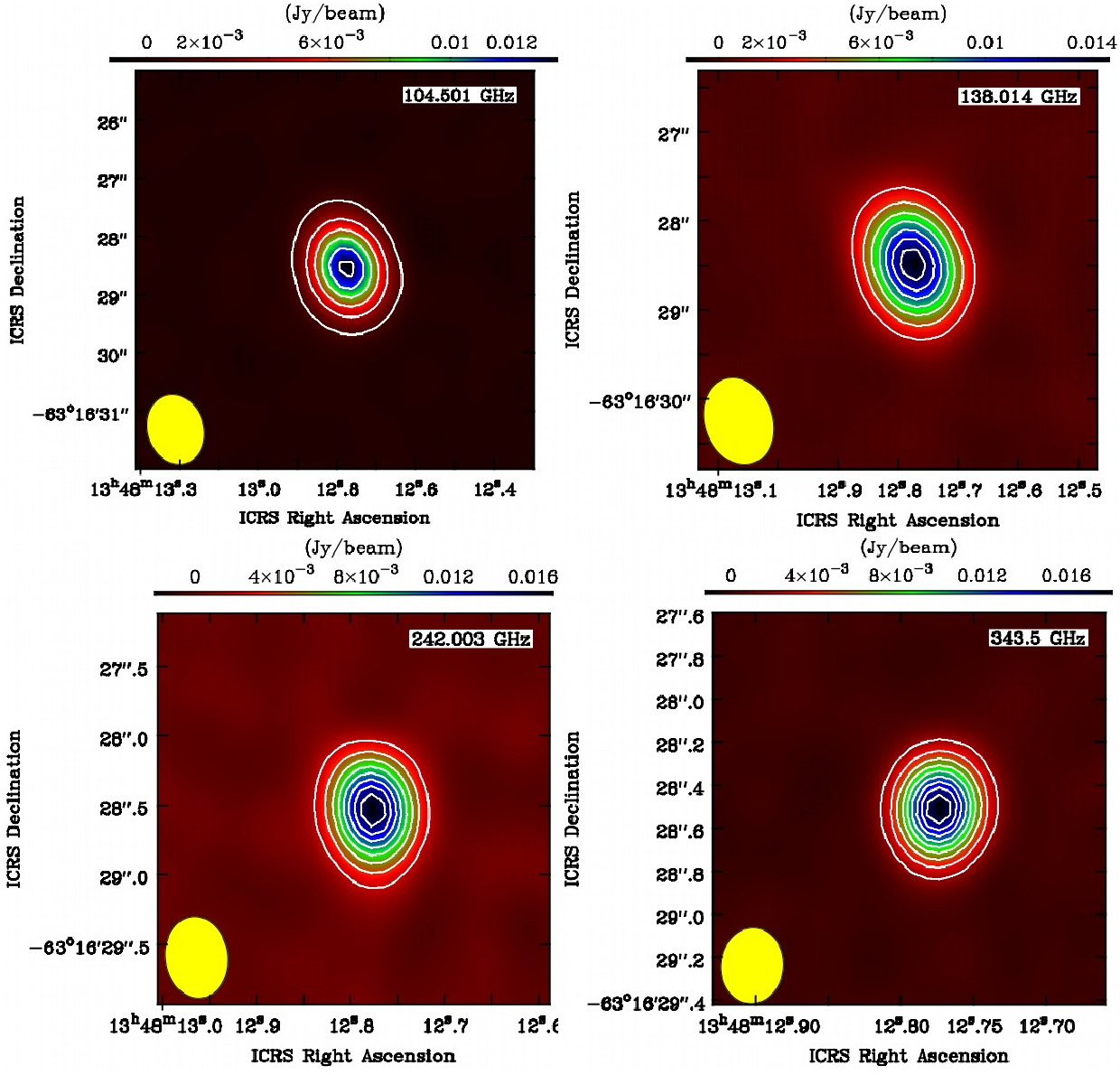}
			\caption{Millimetre-wavelength radio continuum images of the black hole MAXI J1348--630 using {\it ALMA} are obtained at 104.501 GHz ($\sigma$ = 4.84 $\mu$Jy), 138.014 GHz ($\sigma$ = 3.06 $\mu$Jy), 242.003 GHz ($\sigma$ = 8.36 $\mu$Jy), and 343.500 GHz ($\sigma$ = 7.25 $\mu$Jy) frequencies on May 2, 2019, and computed using natural weight. The synthesised beams for observations at each frequency are shown in one corner of each image (shown in yellow). The contour levels start at 3$\sigma$, which is increased by a factor of $\surd$2.}}
			\label{fig:almaimg}
	\end{figure*}

	\section{{Observation and data analysis}}
	\label{sec:obs}
We have used data of \textit{Swift}/BAT (15--50~keV) and \textit{MAXI}/GSC (2--20~keV) to study the temporal variation of flux during the outbursts. The evolution of HR was studied by using \textit{MAXI}/GSC fluxes in the energy
ranges of 2--6~keV and 6--20~keV.
	
\subsection{{{\it ALMA} observation}}
\label{subsec:ALMA}
To study the X-ray binary transient MAXI J1348--630 in radio wavelength,
we have used 12~m array data from the high-resolution Atacama Large Millimeter/Submillimeter Array \textit{ALMA}.\footnote{\href{https://www.almaobservatory.org/en/home/}{https://www.almaobservatory.org/en/home/}.}
The observation log is given in Table~\ref{tab:log}. The black hole was observed in bands 3, 4, 6, and 7 with XX and YY signal correlators. The on-source total integration times during the observation were 302.400~s, 302.400~s, 423.360~s, and 1149.120~s for \textit{ALMA} bands 3, 4, 6, and 7 respectively. The bandwidth of all science spectral windows is 1.875~GHz, which is divided into 1920 channels. During the observation, a total of 43 antennas were working with a minimum baseline of 15.1~m and a maximum baseline of 783.5~m. The antenna temperature did not exceed 100~K. During the observation of MAXI J1348--630, the precipitable water vapour (PWV) of the \textit{ALMA} bands 3, 4, 6, and 7 were 1.1~mm, 1.1~mm, 1.2~mm, and 1.1~mm, which indicates that the environmental condition was good. The phase RMS during the observation of \textit{ALMA} bands 3, 4, 6, and 7 were 0.105$^{\circ}$, 0.163$^{\circ}$, 0.239$^{\circ}$, 0.770$^{\circ}$. During the observation, the source J1427--4206 was used as a bandpass and flux calibrator, and J1337--6503 was used as a phase calibrator.

We used the Common Astronomy Software Application (\texttt{CASA v5.4.1})\footnote{\href{https://casa.nrao.edu/}{https://casa.nrao.edu/}.}
for data reduction and continuum image of black hole MAXI J1348--630 with the standard data reduction pipeline delivered by \textit{ALMA} observatory. The continuum flux density of the flux calibrator J1427--4206 was scaled for each baseline and matched with the Perley-Butler 2017 flux calibrator model with 5\% accuracy using task \texttt{SETJY} \citep{Per07}. After the flux and bandpass calibration, we applied the \texttt{CASA} pipeline with tasks \texttt{hifa\_bandpassflag} and \texttt{hifa\_flagdata} for flagging the bad data. After the initial data calibration, we used the task \texttt{MSTRANSFORM} to split the corrected target data set with all available rest frequencies. After the splitting of target data, we created the continuum image of MAXI J1348--630 using task \texttt{TCLEAN} with Stokes-I and robust parameter 0.5. The continuum image was deconvolved several times to reduce the background RMS. After the initial data calibration, the observed background RMS varied between 15 and 22 mJy beam$^{-1}$ for \textit{ALMA} bands 3, 4, 6, and 7. For improvement of the background RMS, we applied the self-calibration method using the \texttt{CASA} tasks \texttt{GAINCAL} and \texttt{APPLYCAL}. After the self-calibration, the background RMS of the source for \textit{ALMA} bands 3, 4, 6, and 7 varied between 3.06--8.36 mJy beam$^{-1}$. After the creation of the continuum emission map, we applied the task \texttt{IMPBCOR} for the correction of the primary beam pattern.

	\begin{figure}[t]
		\centering{
			\includegraphics[width=5.8cm, angle=270]{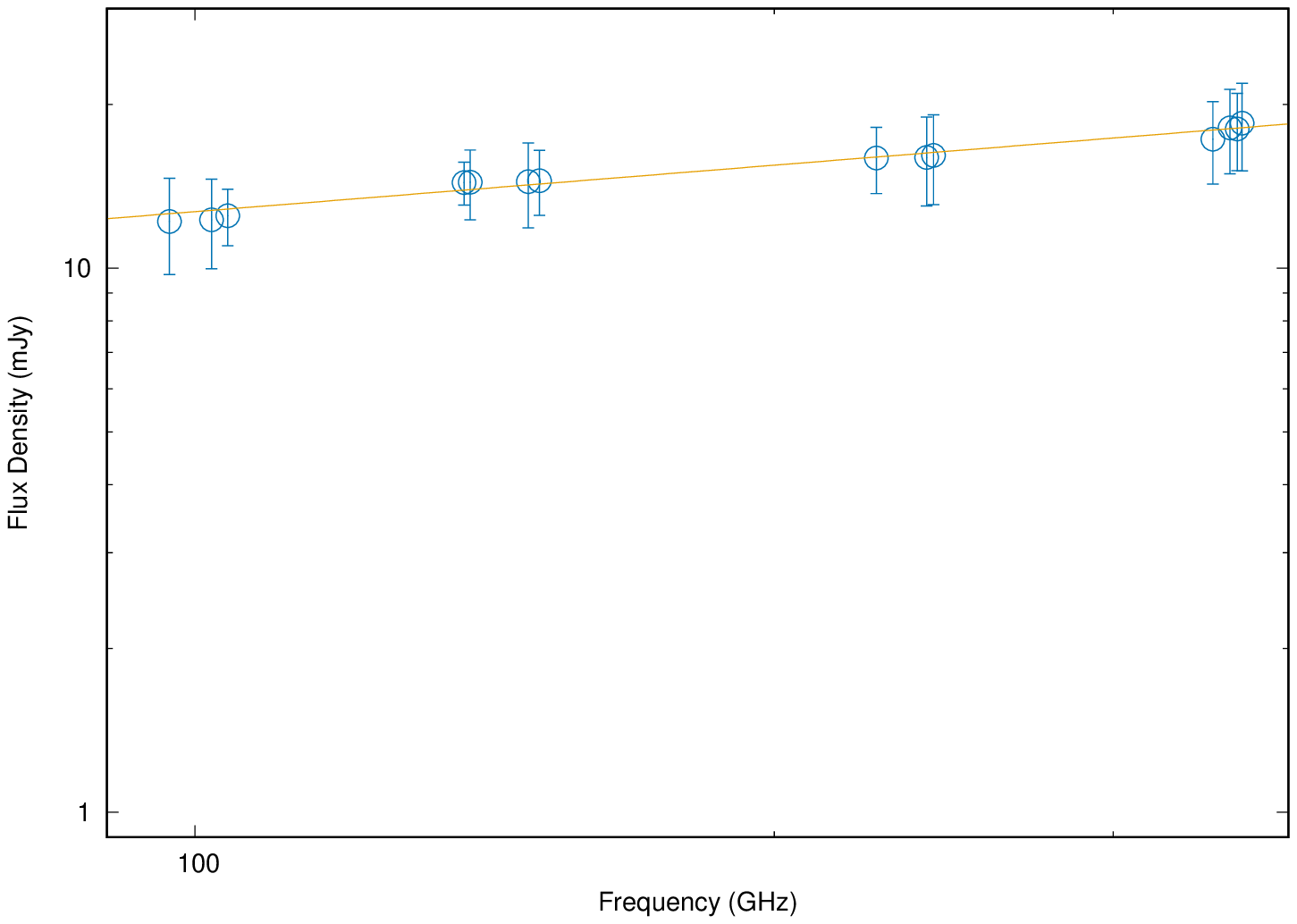}
			\caption{{The radio continuum spectrum of MAXI J1348--630 is in the frequency range from 89.56 GHz to 351.44 GHz, with a variation of flux density between 12.18 mJy and 18.47 mJy. The continuum radio spectrum is fitted using a power law ($F_\nu\propto\nu^{\alpha}$) and the spectral index is found to be $\alpha = 0.28\pm0.02$.}}
			\label{fig:contspec}}
	\end{figure}

\subsection{{{\it NICER} observation}}
\label{subsec:nicer}
The Neutron Star Interior Composition Explorer (\textit{NICER}) on the International Space Station (ISS) is mainly dedicated to the study of neutron stars. The primary instrument of \textit{NICER} is the X-ray Timing Instrument (XTI), which is an array of 56 X-ray photon detectors that collect photons in
the energy range of 0.2--12~keV \citep{Ge16}. \textit{NICER} performed multiple
observations of MAXI J1348--630 after its discovery. We used a quasi-simultaneous
\textit{NICER} observation with \textit{ALMA} during the decay phase of the January 2019 outburst. The observation details are summarised in Table~\ref{tab:log}.

We have used the \texttt{nicerl2} script to process the \textit{NICER} data in
\texttt{HEASOFT} 6.31 with the latest caldb version\break  20221001. We used \texttt{nicerl3-spect} to extract the spectrum from the clean event file. The good time intervals were selected for the timing analysis according to the following criteria: The ISS was not in the South Atlantic Anomaly (SAA) zone, the source elevation was $>$ 20$^{\circ}$ above the Earth limb, and the source direction was at least 30$^{\circ}$ from the bright Earth. The background corresponding to each epoch of the observation was simulated by using the \texttt{nibackgen3C50}\footnote{\url{https://heasarc.gsfc.nasa.gov/docs/nicer/tools/nicer_bkg_est_tools.html}.} tool \citep{Re21}. We used \texttt{XSPEC v12.13} \citep{Ar96} for the spectral fitting.

\subsection{{{\it Swift} observation}}
\label{subsec:swift}
The Neil Gehrels \textit{Swift} Observatory is a multi-wavelength observatory
that operates in a wide range of wavelengths, including optical, ultraviolet, X-ray, and soft gamma-ray wavebands, with three different instruments onboard. The Burst Alert Telescope (BAT) operates in the energy range of 15--150~keV, which detects new events and the corresponding location in the sky \citep{Ba05}. We used daily monitoring data of \textit{Swift}/BAT (15--50~keV) to plot the light curve. The X-ray Telescope (XRT) onboard \textit{Swift} operates in the energy range of 0.3--10~keV to study different spectral properties of a source \citep{Bu00}.

The Ultraviolet/Optical Telescope (UVOT) operates in both UV and visible wavelengths with the help of six different filters (1928--5468 \AA )
\citep{Po08}. We used UVOT observation performed on MJD 58606 close to our quasi-simultaneous \textit{NICER} and \textit{ALMA} observation to study the broadband spectrum during the January 2019 outburst. The observation log and a list of filters used during the observations are shown in Table~\ref{tab:log}. We performed \texttt{uvotdetect} for the source location on the UVOT sky image. We performed the routine \texttt{uvotsource} to examine the flux of the source, keeping the background threshold of $5\sigma $. The result is used in the broadband spectrum.
	
	\begin{figure}[t]
		\centering{
			\includegraphics[width=7.75cm]{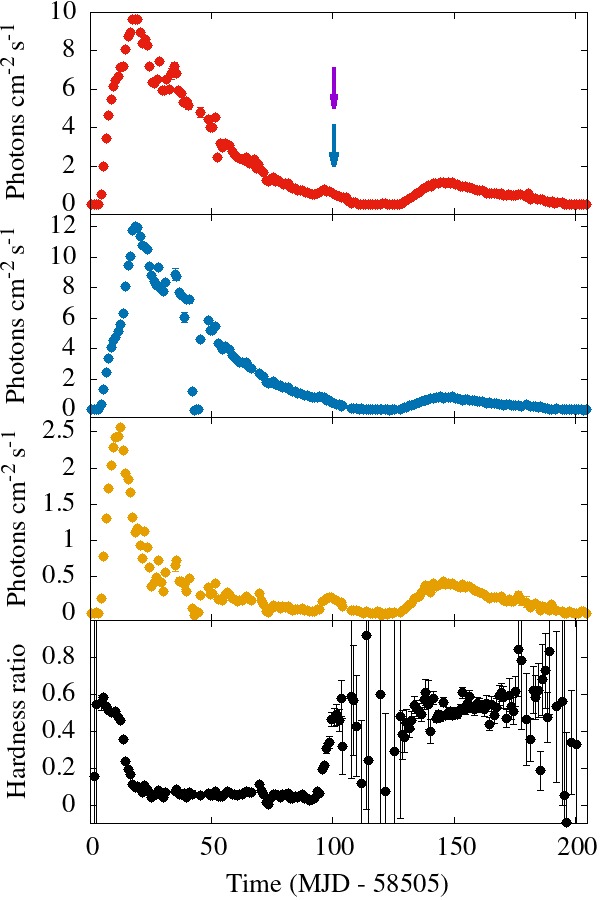}
			\caption{{Variation of flux and hardness ratio towards the source MAXI J1348--630 during the outbursts of January and June 2019 using different energy bands of {\it MAXI}/GSC. The first panel shows the variation of {\it MAXI}/GSC flux in the energy range of 2--20 keV. The second and third panels represent the flux evolution in 2--6 keV and 6--20 keV of {\it MAXI}/GSC respectively. The fourth panel shows the variation of the hardness ratio (6-20 keV/2--6 keV) during the complete evolution of the outbursts. The violet and blue arrows indicate the time of the quasi-simultaneous {\it ALMA} and {\it NICER} observations during the decay phase of the January 2019 outburst.} }
			\label{fig:light}}
	\end{figure}

	\section{{Results}}
	\label{sec:result}
	
\subsection{{Continuum image of MAXI J1348--630 using {\it ALMA}}}
\label{subsec:img}
We study the radio continuum emission from MAXI J1348--630 using \textit{ALMA} band 3, 4, 6, and 7 observations in the frequency range of 89.56~GHz--351.44~GHz. The measured flux densities at different frequencies vary between $\sim $12.18 mJy and $\sim $18.47 mJy. For estimation of the flux densities, we fitted the 2D Gaussian over the black hole candidate MAXI J1348--630 using the \texttt{CASA} task \texttt{IMFIT}. The RMS of the background of the radio continuum images varied between 3.06--8.36 mJy beam$^{-1}$ in different observing frequencies. The continuum images of MAXI J1348--630 at different frequencies are shown in Fig.~\ref{fig:almaimg}. The phase centre of the continuum image of MAXI J1348--630 is RA $=$ 13~h48m12.790s, Dec $=$ --63$^{ \circ}$16$^{\prime}$28.480$^{\prime \prime}$ (J2000). In continuum images, the beam sizes are 1.22$^{\prime \prime}\times $0.97$^{\prime \prime}$ for 104~GHz, 1.22$^{\prime \prime}\times $0.97$^{\prime \prime}$ for 138~GHz, 0.58$^{\prime \prime}\times $0.44$^{\prime \prime}$ for 242~GHz, and 0.35$^{\prime \prime}\times $0.29$^{\prime \prime}$ for 343~GHz. After fitting a 2D Gaussian over the source, we found that the \textit{ALMA} band 3, 6, and 7 images of MAXI J1348--630 did not resolve, but the band 4 image was marginally resolved. So, we cannot draw any conclusions about the continuum morphology of MAXI J1348--630.

The resultant radio continuum spectrum of MAXI J1348--630 is shown in Fig.~\ref{fig:contspec}. The continuum radio spectrum is fitted using power
law ($F_{\nu}\propto \nu ^{\alpha}$) with a spectral index of $\alpha = 0.28\pm 0.02$. The slope of the spectrum is positive in the radio region, indicating that the synchrotron radiation is from an optically thick region and that the self-absorption frequency was more than 351.4~GHz.

We also searched the molecular lines from the surroundings of the source, as found earlier for the cases of galactic black holes GRS 1758--258 and 1E 1740.7--2942 \citep{Te20}. We have looked for molecular line emission of H$_{2}$CO, CH$_{3}$OH, SO$_{2}$, $^{13}$C$^{34}$S, and OH from the background of the source. We cannot find the line emission of those molecules in the background. The derived 3$\sigma $ upper-limit column-density of H$_{2}$CO, CH$_{3}$OH, SO$_{2}$, $^{13}$C$^{34}$S, and OH are $<$1.5$ \times $10$^{12}$ cm$^{-2}$, $<$1.9 $\times $ 10$^{12}$ cm$^{-2}$, $<$3.8$ \times $10$^{13}$ cm$^{-2}$, $<$2.8 $\times $ 10$^{13}$ cm$^{-2}$, and $<$8.5 $\times $ 10$^{13}$ cm$^{-2}$ respectively. This result suggests that there is not enough molecular gas in the vicinity of MAXI J1348--630.

\subsection{{Temporal variations of X-ray emission}}
\label{sec:light_curves}
The Gas Slit Camera (GSC) onboard \textit{MAXI}\footnote{\href{http://maxi.riken.jp/top/index.html}{http://maxi.riken.jp/top/index.html}.} and \textit{NICER} have been continuously monitoring the source since its discovery in January 2019. A subsequent outburst happened in June 2019. In this paper, we mainly focused on the study of the broadband spectrum
during the decay phase of the January outburst using \textit{ALMA}, \textit{Swift}, and \textit{NICER}. The evolution of X-ray flux during the outburst is shown in Fig.~\ref{fig:light} in different energy bands of \textit{MAXI}/GSC. The top panel shows the evolution of 2--20~keV \textit{MAXI}/GSC flux between MJD 58505 and MJD 58710. The second and third panels show the variation of flux in two different energy bands of  \textit{MAXI}/GSC (2--6~keV and 6--20~keV).

The peak flux during the January 2019 outburst was $\sim $20 photons cm$^{-2}$
s$^{-1}$ in the 2--20~keV as measured by \textit{MAXI}/GSC. In the 2--6~keV, the peak flux (12 photons cm$^{-2}$ s$^{-1}$) was observed on MJD 58523, and in the 6--20~keV the peak flux (2.56 photons cm$^{-2}$ s$^{-1}$) was observed on MJD 58517. The bottom panel of Fig.~\ref{fig:light} shows the variation of the hardness ratio (6--20 keV/2--6~keV of \textit{MAXI}/GSC) during these two outburst phases.

\subsection{{Broadband spectrum during the decay phase of January outburst}}
\label{subsec:broadband}
	
\begin{figure}[t]
		\centering{
			\includegraphics[width=5.8cm,angle=270]{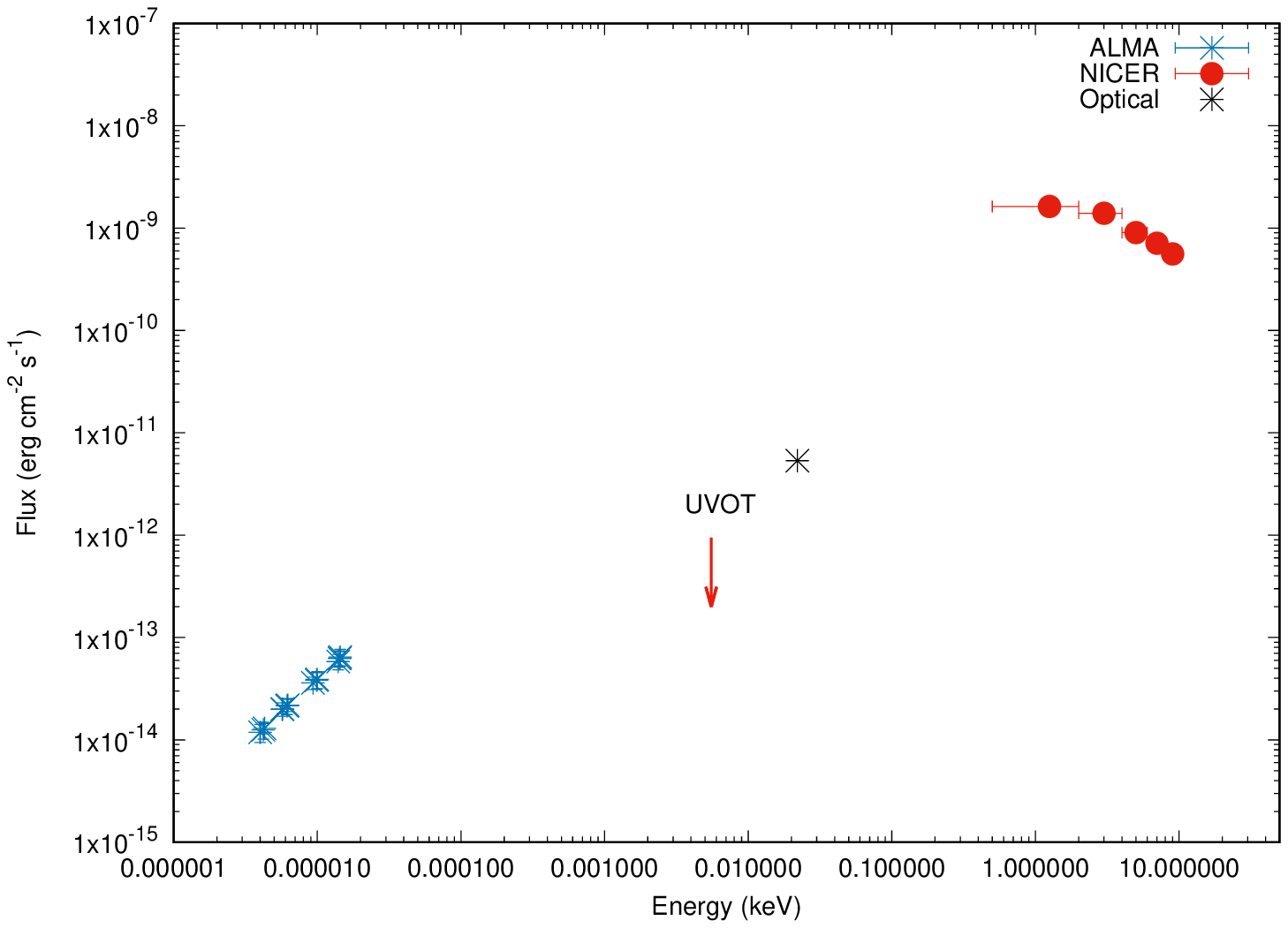}	
			\caption{{Broadband spectrum of the source MAXI J1348--630 from radio to X-ray bands. The {\it NICER} fluxes are shown by red points. The horizontal error bars represent the range for which fluxes are estimated. The errors in flux are also added which are lower than the point size. The {\it NICER} and {\it ALMA} observations were quasi-simultaneous (close to MJD 58605) and we also included the result of the optical observation performed by \citet{De19} on MJD 58509. The down arrow shows the upper limit of {\it Swift}/UVOT flux with the {\it uvm2} filter. 
			}}
			\label{fig:broadband}}
	\end{figure}

We have studied the spectral energy distribution (SED) of MAXI J1348--630
over multi-wavelength regions to get an idea about the emission mechanism
from different emitting regions. The broadband SED has been studied in
a wide range of energies from radio to X-ray, including fluxes at ultraviolet
and optical wavebands during the outburst of January 2019. In Fig.~\ref{fig:broadband}, we have shown the broadband spectrum of the source
using quasi-simultaneous observations. All observations were performed
between May 2 and May 3, 2019, during the decay phase of the January 2019
outburst. \textit{ALMA} observations started at 03:06 UT on May 2, 2019 (MJD
58605.13) from which the radio emission from the compact jets was studied
using different wavebands (bands 3, 4, 6, and 7) with frequencies ranging
from 89.56~GHz to 351.44~GHz. The \textit{NICER} observation performed on
May 2, 2019 (MJD 58605.3) is used for the X-ray flux. Flux in the ultraviolet
region is taken from the \textit{UVOT} observations that were carried out
with \textit{uvm2} filter having a central wavelength ($\lambda $) of 2246
\AA  (details are given in Table~\ref{tab:log}). The source remained undetected
in the \textit{uvm2} band. We have mentioned the $5\sigma $ upper limit of
flux at the location of the source in Fig.~\ref{fig:broadband}.
	
The broadband spectrum is studied when the source is in the intermediate
state, making a transition from SS to HS (time of observations of \textit{ALMA}
and \textit{NICER} are pointed out in Fig.~\ref{fig:light}) in the decay phase of the January 2019 outburst. Emission in the X-ray wavelengths is higher compared to that in other wavelengths, which indicates strong radiation from the accretion disk. In the region of radio wavelengths, the spectrum was nearly flat/slightly inverted. Fitting the \textit{ALMA} spectrum, we obtained the spectral index $(\alpha )=0.28\pm 0.02$ (assuming $F_{\nu}\propto \nu ^{\alpha}$) as measured on May 2, 2019. The spectral index indicates self-absorbing compact jets and a slightly inverted radio spectrum. The low radio flux from MAXI J1348--630 categorized itself as a ``radio-quiet'' black hole candidate.

Due to the lack of sufficient photons in the ultraviolet and optical wavelengths,
we could not fit the broadband spectrum. The X-ray spectrum in the range of 0.5--10~keV obtained from \textit{NICER} can be described well by using the combinational model consisting of a comptonization component \textit{simpl}
\citep{St09} to account for the power law continuum produced by the thermal comptonisation of soft photons in hot gas of electron and a multi-temperature
disk blackbody component \textit{diskbb} \citep{Mit84}, which is used to account for the emission from the optically thick and geometrically thin accretion disk. Typically, in hard and hard-intermediate states of black hole X-ray binaries, the accretion flow consists of an optically thick and geometrically thin outer disk \citep{Sh73}. In addition, the interstellar medium absorption model \textit{tbabs} \citep{Wi00} is used. The photon index during the observation is 1.68 $\pm $ 0.01, indicating the source was close to the hard state, and the hydrogen column density is $\sim $0.3 $\times 10^{22}$ cm$^{-2}$. The inner disk temperature is found to be nearly 0.38 $\pm $ 0.02~keV and the disk normalization is 9796${ ^{+2182}_{-1747}}$. The total unabsorbed flux during the \textit{NICER} observation on MJD 58605.3 was $\sim $5.4 $ \times $ 10$^{-9}$ erg cm$^{-2}$ s$^{-1}$ (1--10~keV). The quasi-simultaneous radio flux densities are found to be 12.2--18.5 mJy (or corresponding fluxes are $1.2 \times 10^{-14}$ -- $6.5 \times 10^{-14}$ erg cm$^{-2}$ s$^{-1}$ respectively) as measured by \textit{ALMA} bands 3, 4, 6, and 7 at frequencies ranging between 97--350~GHz measured during MJD 58605.13--58605.19.

\section{{Radio and X-ray correlation}}
\label{sec:radio_X-ray}
\begin{figure}[t]
		\centering{
			\includegraphics[width=\columnwidth]{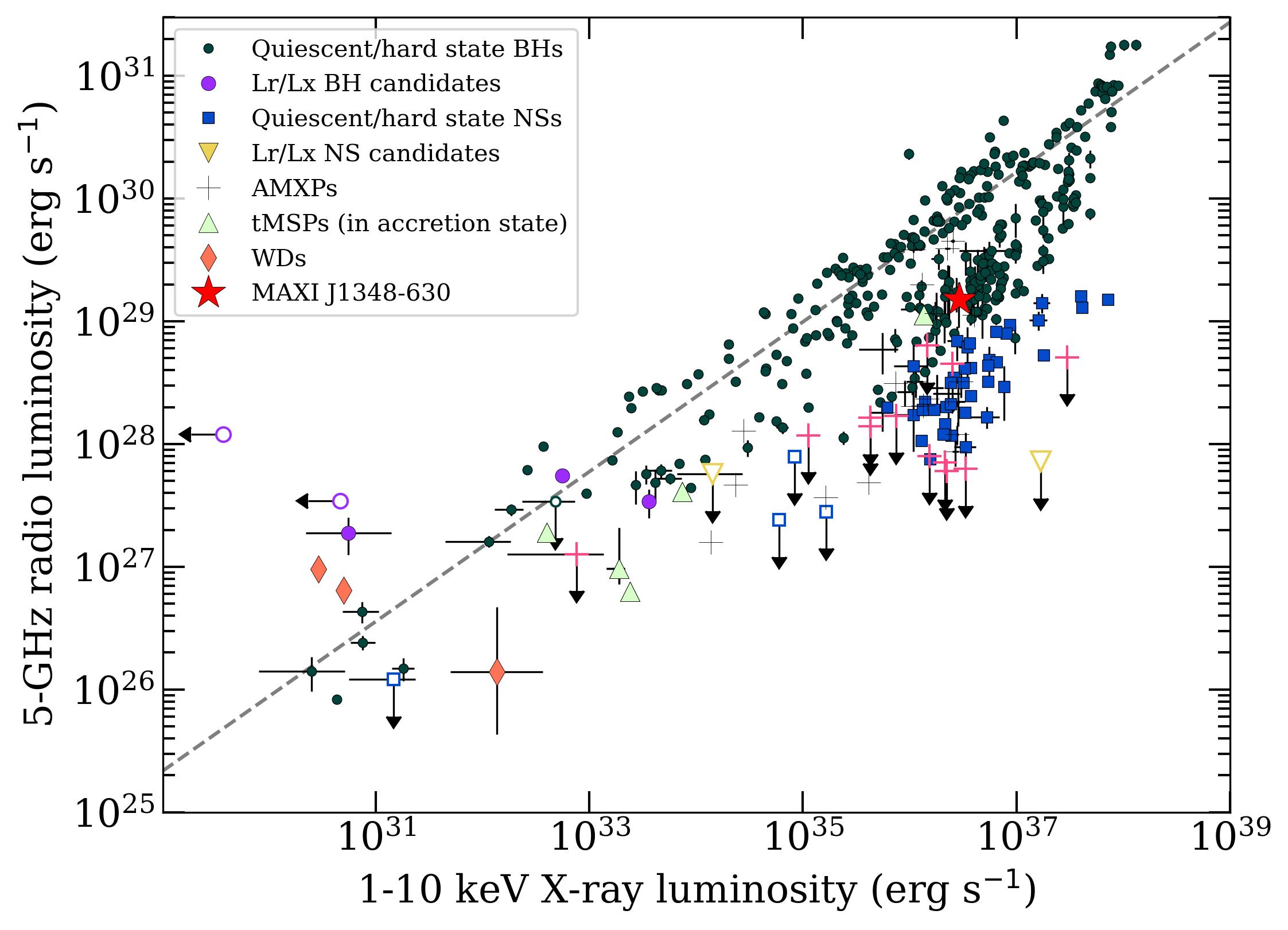}	
			\caption{{Radio/X-ray correlation study of MAXI J1348--630 during the decay phase of the January 2019 outburst using quasi-simultaneous {\it ALMA} and {\it NICER} observations. The figure is adapted from \citet{Bah18}, which shows the radio/X-ray observations for several types of accreting stellar-mass compact objects. MAXI J1348--630 is shown by the red asterisk. The dotted black line indicates the best-fit relation for BHs \citep{Gal06}.}}
			\label{fig:lrlx}}
	\end{figure}

One of the main objectives of this study is to investigate the correlation between the X-ray luminosity and the radio luminosity of MAXI J1348--630 during the quasi-simultaneous observation of \textit{ALMA} (97--350~GHz), \textit{NICER} (1--10~keV), and \textit{Swift}/UVOT. We have estimated the radio luminosity from the radio flux density using \textit{ALMA} data, assuming a source distance of 2.2 kpc \citep{Ch21}. To compare the radio luminosity to other well-known systems we have calculated the corresponding 5~GHz radio flux density from Fig.~\ref{fig:contspec}. The 5~GHz radio flux density is estimated to be $\sim $5 mJy and the corresponding radio luminosity is found to be 1.5 $\times 10^{29}$ erg s$^{-1}$. The X-ray luminosity is also estimated using quasi-simultaneous \textit{NICER} observation on MJD 58605.3. The unabsorbed X-ray flux in the band 1--10~keV is estimated to be 5.4 $\times 10^{-9}$ erg cm$^{-2}$ s$^{-1}$ and the corresponding luminosity is $\sim $3.1 $\times 10^{36}$ erg s$^{-1}$. We have plotted the estimated radio luminosity and corresponding X-ray luminosity in Fig.~\ref{fig:lrlx} to get an idea about the system. Figure~\ref{fig:lrlx} shows the well-known tracks for different black holes and neutron stars in the
$L_{X}$ -- $L_{R}$ diagram. From this figure, it is visible that MAXI J1348--630
is a new member of a restricted group of sources, which is consistent with earlier results \citep{Car21}. \citet{Car21} also used different radio observations using \textit{ATCA} and \textit{MeerKAT} to study the correlation
between the X-ray and radio luminosity and concluded that MAXI J1348--630
did not follow any well-known track in the $L_{X}$ -- $L_{R}$ diagram. During the quiescence, \citet{Ca22} studied the correlation between the X-ray luminosity and radio luminosity and concluded that hybrid-correlation sources like MAXI J1348--630 followed a standard track at low luminosities like GX 339--4, V404 Cyg, and MAXI J1820$+$070 (see Fig.~2 of \citet{Ca22}).

\section{{Discussion}}
\label{sec:discussions}

In this paper, we have mainly focused on the study of the multi-wavelength nature of MAXI J1348--630 during the decay phase of the January 2019 outburst using \textit{ALMA}, \textit{NICER}, and \textit{Swift} data. The correlation between
the X-ray and radio luminosity is performed to understand the nature of the source. During the decay phase of the January 2019 outburst, the source started the soft to hard transition on MJD 58597 and then the source came back to the hard state on MJD 58608 \citep{Zh20}. During the quasi-simultaneous observation of \textit{NICER} and \textit{ALMA} on MJD 58605 the source was close to the hard state. During this state, the \textit{NICER} spectrum is modelled using a disk black body and a comptonization component. We have estimated the ratio between the flux from the comptonization component to the total flux to understand the emission mechanism. The ratio (F$_{\text simpl}$/F$_{ \text Total}$) is found to be $\sim $70\%, which implies that the spectrum is dominated by the comptonised component with a photon index of $\sim $1.7.

In the HS of BHXRBs, the energy spectra are strongly dominated by comptonized emission, with a weak and cool thermal component emitted from truncated disks \citep{Do07,Saha23}. Whereas in SS, emissions are characterized by strong thermal blackbody radiation from the accretion disk, and the effect of comptonization becomes minimum. During this state, the inner disk radius sometimes extends up to the innermost stable circular orbit \citep{Ta89, Eb93, St10}. During the transition towards the hard state, the comptonization component starts dominating the spectra, and fluxes are mostly due to the hard photons \citep{Ta08, Ma08}.

Multi-wavelength observations are key to understanding the accretion mechanism
from different regions of BHXRB during outbursts (viz. \citet{Ud02, Ch05}). While the sources of X-ray and radio emissions are well understood, the origin of emissions in infrared, ultraviolet, and visible wavebands is complicated to understand \citep{va95}, since multiple radiation mechanisms occur in different regions of the system. The soft X-rays are believed to come from the inner-most regions of the accretion disk, whereas the ultraviolet and optical emissions are thought to come from the irradiated outer thin accretion disks \citep{Po14}. The origin of the hard X-rays is due to the inverse comptonization of the disk photons
by the electron clouds around the black hole called ``corona''
\citep{Do07}. Radio emissions originated due to synchrotron emissions from
jets \citep{Fa96, Fa99}.

Earlier, some multi-wavelength observations were performed for other sources to study the accretion phenomenon around them. Earlier, the radio/X-ray diagram of MAXI J1348--630 during quiescence indicated that the source belongs to the standard (radio-loud) track in the quiescence phase \citep{Ca22}. According to typical SEDs of AGNs presented in Fig.~\ref{fig:almaimg} and Fig.~10 of \citet{El94}, the ratio of $\nu I_{\nu}$ between the radio and X-ray band is roughly $10^{-2}$ for radio-loud quasars and $10^{-5}$ for radio-quiet quasars. We have used the radio flux from the 240~GHz \textit{ALMA} band, which is $3.8\times 10^{-14}$ erg cm$^{-2}$ s$^{-1}$ and the unabsorbed \textit{NICER} flux in the 1--10~keV band was $5.4\times 10^{-9}$ erg cm$^{-2}$ s$^{-1}$. We have estimated the ratio of $\nu I_{\nu}$ for MAXI J1348--630 using fluxes in radio (\textit{ALMA} band, $\nu = 10^{11}$ Hz) and X-ray (\textit{NICER} band, $\nu = 10^{18}$ Hz) bands.
%
\begin{equation}
Ratio = \frac{(\nu I_{\nu})_{Radio}}{(\nu I_{\nu})_{X-ray}} =
\frac{2.4\times 10^{11}\times 3.8\times 10^{-14}}{{1.2\times 10^{18}} {\times 5.4\times 10^{-9}}}
\label{eq1}
\end{equation}
The estimated ratio for MAXI J1348--630 is $\sim 1.5\times 10^{-12}$, which is lower than the typical value ($\sim 10^{-5}$) for radio-quiet quasars. This indicates that MAXI J1348--630 falls into the regime of ``radio-quiet''. During the quiescence, \citet{Car22} studied the radio and X-ray correlation and concluded that MAXI J1348--630 falls in the well-defined track for the black holes in the $L_{X}$ -- $L_{R}$ diagram. The results indicated that MAXI J1348--630 belongs to the standard ``radio-loud'' track during the quiescence. Misalignment of jets may be possible for the low radio emission from the source \citep{Mi19} during the outburst. In the multi-wavelength campaign on the black hole X-ray binary GRS 1915$+$105 (April 2002), it was found that radio emissions from jets were relatively weak, whereas the X-rays emitted from the disk were much stronger \citep{Ud02}. For GRS 1915$+$105, the ratio was found to be $\sim 10^{-7}$ \citep{Ud02}. The same nature of SED was observed in the LMXB GX 339--4 during the outburst of 2010 \citep{Ca11}. But during the broadband study of the HMXB SS 433, \citet{Ch05} noticed that the source was strong in radio wavelengths, but fluxes drastically decreased in the infra-red, optical, and ultraviolet regions, and significantly less in X-ray wavelengths.
	
\section{{Conclusions}}
\label{sec:conclusion}
We report the multi-wavelength spectral properties of the black hole X-ray
binary MAXI J1348--630 during the decay phase of the January 2019 outburst
using the quasi-simultaneous data from \textit{ALMA}, \textit{NICER}, and \textit{Swift}. We found that MAXI J1348--630 falls in the ``radio-quiet'' regime during our multi-wavelength observation. We also study the radio/X-ray correlation
for the source using the \textit{ALMA} and \textit{NICER} data during the decay
phase of the outburst. The correlation study suggested that MAXI J1348--630 did not follow any well-known track in the $L_{X}$--$L_{R}$ diagram and it can be categorized as a new candidate of a restricted group of sources. We have found that the ratio of flux between the comptonised component and total flux (F$_{\text simpl}$/F$_{\text Total}$) is $\ge $ 70\%, indicating the source is close to the hard state in the decay phase of the outburst. The \textit{NICER} spectral fitting results suggest that spectra are dominated by the thermally comptonised photons, as usually observed in the hard state of black hole X-ray binaries.

\section*{Acknowledgments}
We thank the reviewer. This research has made use of data provided by the High Energy Astrophysics Science Archive Research Centre (HEASARC), which is a service of the Astrophysics Science Division at NASA/GSFC. We acknowledge the use of public data from the \textit{NICER}, and \textit{swift} data archives. Special thanks to Neil Gehrels \textit{Swift} Observatory and MAXI-RIKEN teams since daily monitoring data of \textit{Swift}/BAT and \textit{MAXI}/GSC have been used in this research. This paper uses the \textit{ALMA} data. \textit{ALMA} is a partnership of ESO (representing its member states), NSF (USA), and NINS (Japan), together with NRC (Canada), MOST and ASIAA (Taiwan), and KASI (Republic of Korea), in cooperation with the Republic of Chile. The Joint \textit{ALMA} Observatory is operated by ESO, AUI/NRAO, and NAOJ. \\\\

\section*{Data availability}
{{The X-ray data used for this article are publicly available in the High
Energy Astrophysics Science Archive Research Centre (HEASARC) at
		\href{https://heasarc.gsfc.nasa.gov/cgi-bin/W3Browse/w3browse.pl}{https://heasarc\\.gsfc.nasa.gov/cgi-bin/W3Browse/w3browse.pl}.
		The {\it ALMA} data are publicly available at \href{https://almascience.nao.ac.jp/asax/}{https://almascience.nao.ac.jp/asax/} (project id: 2018.1.01034.T).	}} 

\section*{Funding} The authors declare that no funds, grants, or other supports were received during the preparation of this manuscript.

\section*{Conflicts of interest}
The authors have no relevant financial or non-financial interests to disclose.

\section*{Author Contributions}
All authors contributed to the conception and design of the study. Data analysis is performed by MM, DS, and AM. The manuscript is jointly written by MM, SP, and DS. All authors read and approved the final manuscript.

\makeatletter
\let\clear@thebibliography@page=\relax
\makeatother

\end{document}